\documentclass[twocolumn]{aastex631}

\usepackage{CJK}
\usepackage{makecell}
\usepackage{multirow}
\usepackage{graphicx}

\newcommand{\teff}{$T_{\rm eff}$}
\newcommand{\logg}{$\log{g}$}
\newcommand{\mh}{[M/H]}

\newcommand{\chitwo}{$\chi^2$}
\newcommand{\halpha}{H$\alpha$}
\newcommand{\hbeta}{H$\beta$}
\newcommand{\hdelta}{H$\delta$}
\newcommand{\hgamma}{H$\gamma$}

\newcommand{\sldelta}{$S_{\rm [l,H\delta]}$}
\newcommand{\slgamma}{$S_{\rm [l,H\gamma]}$}
\newcommand{\slbeta}{$S_{\rm [l,H\beta]}$}
\newcommand{\slalpha}{$S_{\rm [l,H\alpha]}$}
\newcommand{\slm}{$S_{\rm [l,metal]}$}

\newcommand{\smb}{$S_{\rm [m,b]}$}
\newcommand{\smr}{$S_{\rm [m,r]}$}
\newcommand{\smalpha}{$S_{\rm [m,H\alpha]}$}

\newcommand{\rv}{$RV$}
\newcommand{\rvlm}{$RV_{\rm [l,m]}$}
\newcommand{\rvlmetal}{$RV_{\rm [l,metal]}$}

\newcommand{\rvldelta}{$RV_{\rm [l,H\delta]}$}
\newcommand{\rvlgamma}{$RV_{\rm [l,H\gamma]}$}
\newcommand{\rvlbeta}{$RV_{\rm [l,H\beta]}$}
\newcommand{\rvlalpha}{$RV_{\rm [l,H\alpha]}$}

\newcommand{\rvmm}{$RV_{\rm [m,metal]}$}

\newcommand{\rvmalpha}{$RV_{\rm [m,H\alpha]}$}

\newcommand{\rvmetal}{$RV_{\rm metal}$}
\newcommand{\rvalpha}{$RV_{\rm H\alpha}$}

\newcommand{\kms}{km/s}
\newcommand{\aptp}{$A_\mathrm{ptp}$}

\begin{document}
\begin{CJK*}{UTF8}{gbsn}
\title{Stellar atmospheric parameters of $\sim$ 11,000 RR Lyrae stars from LAMOST Spectra}

\correspondingauthor{Jianrong Shi}
\email{sjr@nao.cas.cn}

\author{Jiangtao Wang}
\affiliation{CAS Key Laboratory of Optical Astronomy, National Astronomical Observatories, Beijing 100101, People’s Republic of China}

\author{Jianrong Shi}
\affiliation{CAS Key Laboratory of Optical Astronomy, National Astronomical Observatories, Beijing 100101, People’s Republic of China}
\affiliation{School of Astronomy and Space Science, University of Chinese Academy of Sciences, Beijing 100101, People's Republic of China}

\author{Jianning Fu}
\affiliation{Institute for frontiers in astronomy and astrophysics, Beijing Normal University, Beijing 102206, China}
\affiliation{Department of Astronomy, Beijing Normal University, Beijing 100875, People’s Republic of China}

\author{Weikai Zong}
\affiliation{Institute for frontiers in astronomy and astrophysics, Beijing Normal University, Beijing 102206, China}
\affiliation{Department of Astronomy, Beijing Normal University, Beijing 100875, People’s Republic of China}

\author{Chunqian Li}
\affiliation{CAS Key Laboratory of Optical Astronomy, National Astronomical Observatories, Beijing 100101, People’s Republic of China}
\affiliation{School of Astronomy and Space Science, University of Chinese Academy of Sciences, Beijing 100101, People's Republic of China}



\begin{abstract}

Accurate determination of the stellar atmospheric parameters of RR Lyrae stars (RRLs) requires short individual exposures of the spectra to mitigate pulsation effects.
We present improved template matching methods to determine the stellar atmospheric parameters of RRLs from single-epoch spectra of LAMOST (Large Sky Area Multi-Object
Fiber Spectroscopic Telescope, also known as the Guoshoujing telescope).
We determine the radial velocities and stellar atmospheric parameters (effective temperature: \teff, surface gravity: \logg, and metallicity: \mh) of 10,486 and 1,027 RRLs from 42,729 low-resolution spectra (LRS) and 7,064 medium-resolution spectra (MRS) of LAMOST, respectively.
Our results are in good agreement with the parameters of other databases, where the external uncertainties of \teff, \logg, and \mh\ for LRS/MRS are estimated to be 314/274\,K, 0.42/0.29\,dex, and 0.39/0.31\,dex, respectively.
We conclude with the variation characteristics of the radial velocities (\rv) and stellar atmospheric parameters for RRLs during the pulsation phase.
There is a significant difference of $28\pm21$\,km/s between the peak-to-peak amplitude (\aptp) of \rv\ from \halpha\ line (\rvalpha) and from metal lines (\rvmetal) for RRab, whereas it is only $4\pm17$\,km/s for RRc.
The \aptp\ of \teff\ is $930\pm456$ and $409\pm375$\,K for RRab and RRc, respectively.
The \logg\ of RRab show mild variation of approximately $0.23\pm0.42$\,dex near the phase of $\varphi = 0.9$, while that of RRc almost remains constant.
The \mh\ of RRab and RRc show a minor variation of about $0.25\pm0.50$ and $0.28\pm0.55$\,dex, respectively, near the phase of $\varphi = 0.9$.
We expect that the determined stellar atmospheric parameters would shed new light on the study of stellar evolution and pulsation, the structure of the Milky Way, as well as other research fields.
\end{abstract}



\section{Introduction} \label{sec:intro}

The RR Lyrae stars (RRLs) are low-mass, old (with ages exceeding 10 Gyr), and generically metal-poor \citep{1989PASP..101..570W, 2020A&A...641A..96S, 2022ApJ...931..131M}. 
They reside at the intersection of the Classical Instability Strip (IS) and the horizontal branch on the Hertzsprung-Russell diagram (HRD), having evolved to stages with both helium core and hydrogen shell burning.
Their pulsations are driven by the $\kappa$ mechanism, which operates in the He\,II partial ionization regions \citep{1973ApJ...184..201C, 1992MNRAS.255P...1K}. 
RRLs are large-amplitude variables with periods ranging from about 0.2 to 1.0 days and pulsations amplitudes of about $0^m.2$ to $2^m.0$ in the visual band.
Based on the pulsation modes, RRLs are classified as RRab for the radial fundamental mode, RRc for the first overtone radial mode, and RRd when both modes are present \citep{2011AcA....61..285S}.

Since RRLs have high brightness \citep[$M_{\rm V} \sim 0.65$\,mag,][]{2018MNRAS.481.1195M} and distinctive light curves, they can be easily identified.
A large number of RRLs have been discovered through time-domain photometric surveys in recent years.
In particular, the high-precision light curves provided by \textit{Kepler} \citep{2010ApJ...713L..79K} have revealed many new and unpredictable pulsation phenomena in RRLs \citep{2010ApJ...713L.198K, 2011MNRAS.411..878K}. 
For example, the discovery of low-amplitude modes and period-doubling has enabled the study of nonlinear and non-radial astroseismology \citep{2012ApJ...757L..13M, 2022MNRAS.515.3439N}, providing new insights into the physical mechanisms of the century-old mysterious Blazhko effect \citep{2010MNRAS.409.1244S, 2011MNRAS.414.1111K}, which is the periodic modulation of amplitude and/or phase in the light curves\citep{1907AN....175..325B}.

The spectral observation of RRLs is quite limited compared with the photometry surveys, and the radial velocities and stellar atmospheric parameters determined through spectroscopy are invaluable for investigating the structure, chemical and kinematic properties of the Milky Way \citep{2017ApJ...846...10A, 2022MNRAS.513.1958W, 2022MNRAS.517.2787L}.
However, the determination of stellar atmospheric parameters for RRLs through spectra is a challenging task.

The large amplitude of expansion and contraction of RRLs lead to very drastic variations in the stellar physical parameters during the pulsation cycle, especially in effective temperature \citep{1995ApJS...99..263B, 2011ApJS..197...29F}.
Consequently, large-aperture telescopes are required to obtain high signal-to-noise (SNR) spectra for short exposures to weaken the pulsation effect, resulting in high-resolution spectra (HRS) for only a small number of RRLs.
For example, \citet{2021ApJ...914...10C} and \citet{2023A&A...678A.138K} have collected only 162 and 269 samples of RRLs with HRS, respectively.
Although the metallicity can be estimated from Fourier parameters of the light curves 
\citep{1996A&A...312..111J, 2021ApJ...912..144M} and the $\Delta$S method from the low-resolution spectrum \citep{1959ApJ...130..507P, 2012Ap&SS.341...89W}, there is still a lack of the determination of \teff\ and \logg.

In recent years, a multitude of extensive spectroscopic surveys have been initiated, including RAVE \citep[Radial Velocity Experiment,][]{2003ASPC..298..381S, 2020AJ....160...83S}, 
BRAVA \citep[The Bulge Radial Velocity Assay,][]{2007ApJ...658L..29R, 2020AJ....159..270K},
SEGUE \citep[Sloan Extension for Galactic Understanding and Exploration,][]{2009AJ....137.4377Y}, 
APOGEE \citep[Apache Point Observatory Galactic Evolution Experiment,][]{2017AJ....154...94M, 2022ApJS..259...35A}, 
GALAH \citep[GALactic Archaeology with HERMES,][]{2015MNRAS.449.2604D, 2021MNRAS.506..150B}, and 
LAMOST \citep[Large Sky Area Multi-Object Fiber Spectroscopic Telescope, also known as the Guoshoujing telescope,][]{2012RAA....12.1197C}.
These surveys provide a valuable opportunity to obtain the stellar atmospheric parameters for a large sample of RRLs.
However, a significant limitation arises when the parameter is derived from co-added spectra taken across multiple individual exposures.
This method can introduce phase contamination, particularly when observations occur over different phases \citep{2011ApJS..197...29F}.
Also, the intrinsic swift variations in the atmospheres of RRLs challenge conventional methods, which typically assume static atmospheres \citep{2010A&A...519A..64K}.
This is particularly evident in the distorted line profiles observed at the peak luminosity phase \citep{2019A&A...623A.109G}. 
For instance, as shock waves traverse the atmosphere during specific phases, they can induce broadening, doubling, or even the disappearance of \halpha\ or metal lines in the majority of RRab \citep{2009A&A...507.1621P, 2014A&A...565A..73G,2014NewA...26...72Y, 2021MNRAS.500.2554B, 2022AJ....163..109P}. 
Such phenomena are also noted, albeit less frequently, in RRc and RRd \citep{2021ApJ...909...25D}.

The LAMOST is capable of performing simultaneous spectroscopic observations of 4,000 targets \citep{2012RAA....12.1197C}.
The LAMOST low-resolution regular survey, which started in October 2012, classified observations into V-, B-, M-, and F-plate modes based on the target magnitude, with exposure times of 10, 25, 30, and 30 minutes respectively, as detailed in \citet{2015RAA....15.1095L}.
In September 2018, LAMOST began the medium-resolution survey during bright nights, which adopts time-domain (TD) and non-time-domain (NT) modes for distinct scientific goals, with a uniform exposure duration of 20 minutes \citep{2020arXiv200507210L}.
For instance, the second phase of the LAMOST-$Kepler/K2$ project aims for at least 60 observations covering 50,000 interesting stars within the $Kepler$ and $K$2 fields \citep{2020ApJS..251...15Z}.
At present, LAMOST DR10\footnote{www.lamost.org} has released more than 20 million spectra. Therefore, we employ single-epoch of LRS and MRS from LAMOST to determine the stellar atmospheric parameters of RRLs.

The structure of the paper is as follows. 
Section\,\ref{sec:odata} presents a sample of RRLs collected from multiple photometric survey projects and cross-matches these with LAMOST spectra. 
The methodology used to determine the stellar atmospheric parameters of RRLs from LAMOST spectra is introduced in Section\,\ref{sec:meth}. 
In Section\,\ref{sec:error}, we analyze the internal and external uncertainty of the radial velocities and stellar atmospheric parameters, as well as present the variations characteristics of these parameters during the pulsation cycle.
The properties of stellar atmospheric parameters of RRLs and their application prospects are discussed in Section\,\ref{sec:disc}.
A brief summary is provided in Section \ref{sec:summ}.

\section{Sources and Data}
\label{sec:odata}

\subsection{RR Lyrae sample} 
\label{subsec:samp}

Over the past decade, a large number of RRLs have been found from the explosive development of TD photometry surveys. 
Considering the areas of the LAMOST observations (Dec. $\geq-10\degr$), we collect RRLs samples from the following four surveys.
\begin{itemize}
\item 
$Gaia$: The $Gaia$ space telescope, launched successfully in December 2013, aims to provide accurate positional, photometric, and astrometric parameters for over a billion stars \citep{2016A&A...595A...1G, 2021A&A...649A...1G}.
The RRLs catalog of $Gaia$ DR3 contains 270,905 RRLs, classified through the $Gaia$ DR3 multi-band ($G$, $BP$, and $RP$ band) light and radial velocity curves using the SOS Cep\&RRL pipeline \citep{2022arXiv220606278C}.

\item 
ASAS-SN: The ASAS-SN \citep[The All-Sky Automated Survey for Supernovae,][]{ 2018MNRAS.477.3145J} is an all-sky optical survey project, that has confirmed about 1.5 million variable stars to date, of which $\sim$ 45,065 are RRLs \citep{2023MNRAS.519.5271C}. 

\item 
ZTF: The ZTF \citep[The Zwicky Transient Facility,][]{2019PASP..131f8003B} is an optical survey project focused on the Northern sky. 
Its primary scientific objectives include the study of transient objects, stellar variability, and solar system science. 
A total of 46,393 RRLs have been classified using the light curves from ZTF DR2 \citep{2020ApJS..249...18C}.
\item 
PS1: The PS1 \citep[The Pan-STARRS1 survey,][]{2010SPIE.7733E..0EK, 2016arXiv161205560C} is a multi-epoch and multi-color photometry project with a wide field of view. 
The PS1 3$\pi$ survey is the largest sub-project within it, observing the Northern sky with declinations greater than -30\textdegree. 
Approximately 239,044 RRLs candidates are obtained using data from the PS1 3$\pi$ survey. \citep{2017AJ....153..204S}.
\end{itemize}

\begin{figure*}
\centering
\includegraphics[width=\textwidth]{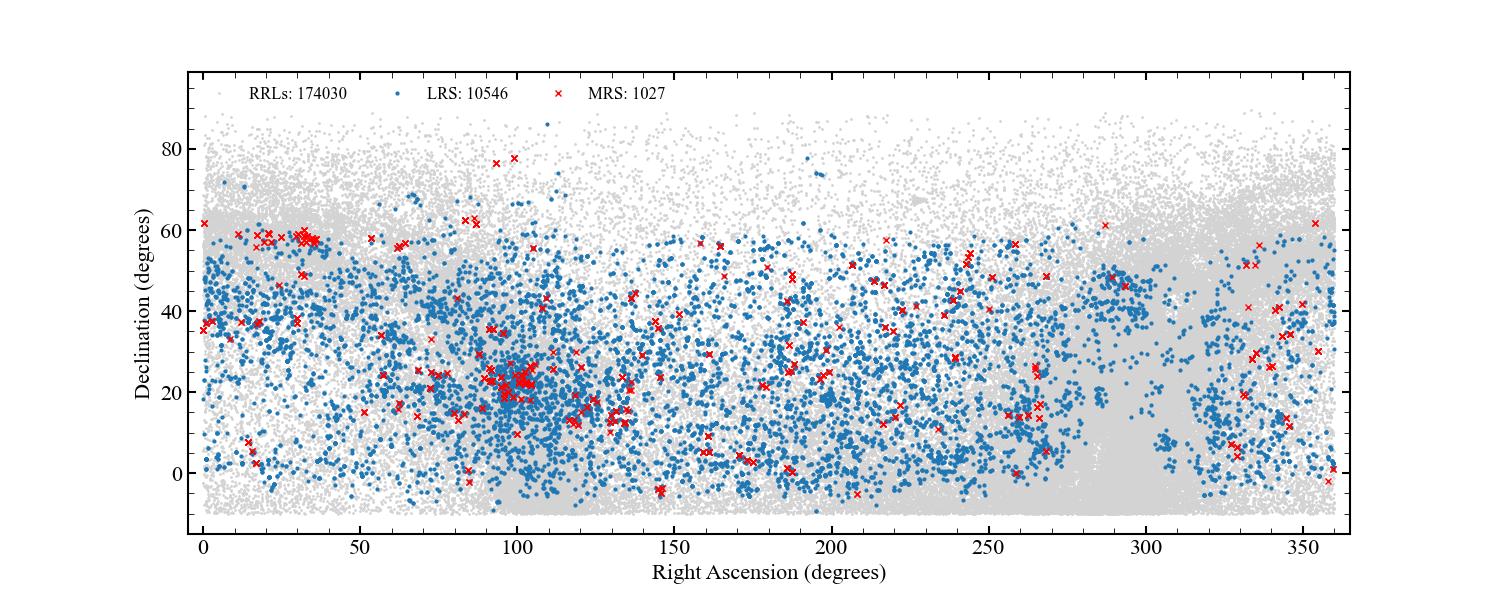}
\caption{The sky coverage of RRLs sample.
The gray dots represent the RRLs sample from five surveys located in the LAMOST field (Dec. $\geq-10\degr$).
The blue dots and red crosse represent RRLs sample with LAMOST LRS and MRS, respectively.
\label{fig:rrls_sky}}
\end{figure*}

Although these four surveys provide a large sample of RRLs, there is an overlap in targets between those catalogs.
Therefore, we cross-matched these catalogs with $Gaia$ DR3, and then obtained the total RRLs sample without overlap by the $Gaiadr3.gaia\_source$ ($Gaiaid$).
The criterion for cross-matching is to select the smallest coordinate separation ($\Delta d$) for the same target within the range where $\Delta d \leq 1.0$ arcsec.
We have collected about 449,093 unique RRLs candidates, of which 174,030 are located within the LAMOST field.
We prioritize the completeness of the sample over the purity of the RRLs sample, so that more RRLs spectra can be obtained from LAMOST.
Thus, the number of RRLs candidates is not constrained by the probability of classification from the four catalogs.
The probabilities from the four catalogs for the sample of RRLs are provided in our results.
Figure \ref{fig:rrls_sky} illustrates the sky position of the 174,030 RRLs (gray dots).

\subsection{LAMOST Spectroscopy}
\label{subsec:data}

The LRS (Resolution, $R\sim 1800$) published by LAMOST is a one-dimensional spectrum with wavelength- and flux-calibration, covering the wavelength range of ($3700 \leq \lambda \leq 9000$\,\AA), and is co-added from the spectra observed at the same night \citep{2015RAA....15.1095L}.
Therefore, we chose the single-epoch LRS, which is processed through the LAMOST 2D pipeline, as detailed in \citet{2021RAA....21..249B}.
The single-epoch LRS is a one-dimensional wavelength-calibrated spectrum containing the blue- ($3700 \leq \lambda \leq 5900$\,\AA) and red-arms ($5700 \leq \lambda \leq 9000$\,\AA).
The single-epoch MRS ($R\sim7500$) contains the blue- and red-arm with wavelength range covers of $4950 \leq \lambda \leq 5350$\,\AA\ and $6300 \leq \lambda \leq 6800$\,\AA, respectively \citep{2018SPIE10707E..2BL, 2020arXiv200507210L}.

We employ the TOPCAT software \citep{2005ASPC..347...29T} to cross-match the coordinates between the RRL sample from survey projects and the LRS and MRS datasets from LAMOST DR10. 
The criterion for overlap is based on the closest target with a coordinate separation of $\Delta d \leq 3.7$ arcseconds
\citep{2020ApJS..251...27W} and SNR $\geq$ 10.
We identify potential contaminants for the RRLs spectra within a range of $\Delta d \leq 3.7$ arcseconds based on the criterion of $G_{\rm other} \leq G_{\rm RRLs} - 1$, where $G_{\rm other}$ and $G_{\rm RRLs}$ represent the magnitudes of potential contaminating targets and the RRLs sample, respectively.
The number of potential contaminants is provided in our results.

\begin{figure*}
\centering
\includegraphics[width=\textwidth]{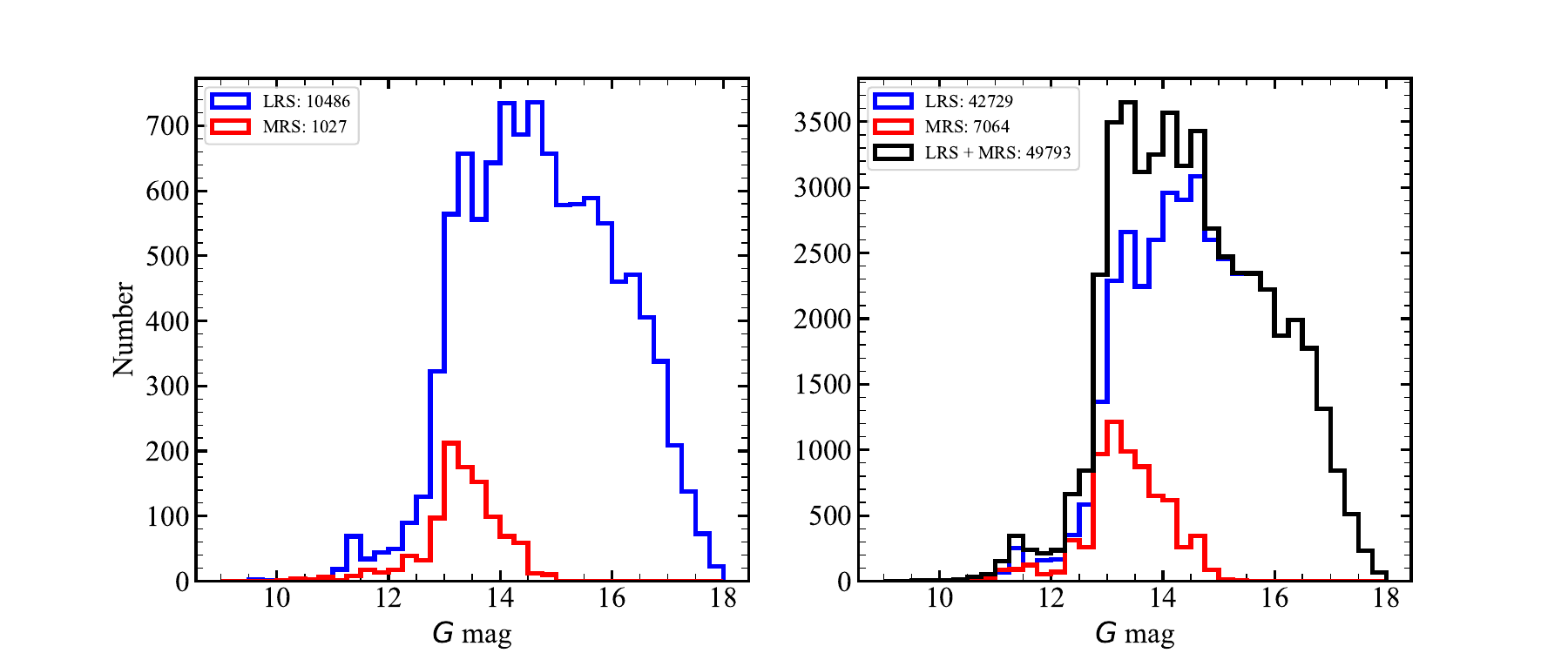}
\caption{The $G$ magnitudes distribution with a bin width of $0.2^m$ for our RRLs sample and their LAMOST spectra.
The right panel shows the LAMOST spectra and the left panel shows our RRLs sample.
\label{fig:rrls_hist}}
\end{figure*}

The position in the sky for RRLs sample with LRS and MRS is shown in Figure\,\ref{fig:rrls_sky}.
The distribution of $Gaia$ $G$ band magnitude for our 10,883 RRLs sample, along with their 49,793 LAMOST spectra, is illustrated in Figure \ref{fig:rrls_hist}.
The $G$ magnitude of the 42,729 LRS for 10,486 RRLs is mainly distributed in $G\sim 13^m - 18^m$, while that of the 7,064 MRS of 1,027 RRLs is mainly distributed in $G\sim 13^m - 15^m$.

\section{Methodology}
\label{sec:meth}

\subsection{Synthetic spectra}
\label{subsec:temp}

The templates are created with three free stellar physical parameters, i.e. \teff, \logg, and \mh, and the ranges and intervals of these parameters are shown in the first two columns of Table\,\ref{tab:sapgrid}.
The synthetic spectra are generated by the iSpec software of the python framework \citep{2014ascl.soft09006B}, where the calculations are performed using the SPECTRUM code \citep{1994AJ....107..742G} with the Kurucz stellar atmosphere models \citep{2005MSAIS...8...14K}, the atomic line-lists of the Vienna Atomic Line Database \citep[VALD3,][]{2011BaltA..20..503K} and the solar abundances from \citet{2009ARA&A..47..481A} have been adopted.
We assume a limb-darkening coefficient of 0.6, and the enhancement of the $\alpha$ element is calculated using Equation 7.12 from \citet{2008oasp.book.....G}.
We calculated 5824 synthetic spectra for each resolution of $R = 1800$ and $R = 7500$.
The last two columns of Table\,\ref{tab:sapgrid} are the resolution and wavelength range of synthetic spectra for the LAMOST LRS and MRS, respectively.

\begin{deluxetable*}{lcc|c|c}[ht]
\label{tab:sapgrid}
\tablecaption{The properties of the template library.}
\tablehead{
\colhead{Parameters} & 
\colhead{range} & 
\colhead{step}  &
\colhead{LRS}  &
\colhead{MRS}}
\startdata
\teff\ (K)    & 5500 - 8000 & 100 & \multirow{3}{*}{\makecell[c]{$R = 1800$ \\ $3600 \leq \lambda \leq 9100$ \AA}} & 
\multirow{3}{*}{\makecell[c]{$R = 7500$ \\ $4850 \leq \lambda \leq 5450$ \AA \\ $6200 \leq \lambda \leq 6900$ \AA}}\\
\logg\ (dex)  & 1.4 - 4.0   & 0.2 & &  \\
\mh\ (dex)    & -3.0 - 0.0  & 0.2 & &  \\
\enddata
\end{deluxetable*}

\subsection{Spectra reduce and Radial velocities}
\label{subsec:specred}

\begin{figure*}
\centering
\includegraphics[width=\textwidth]{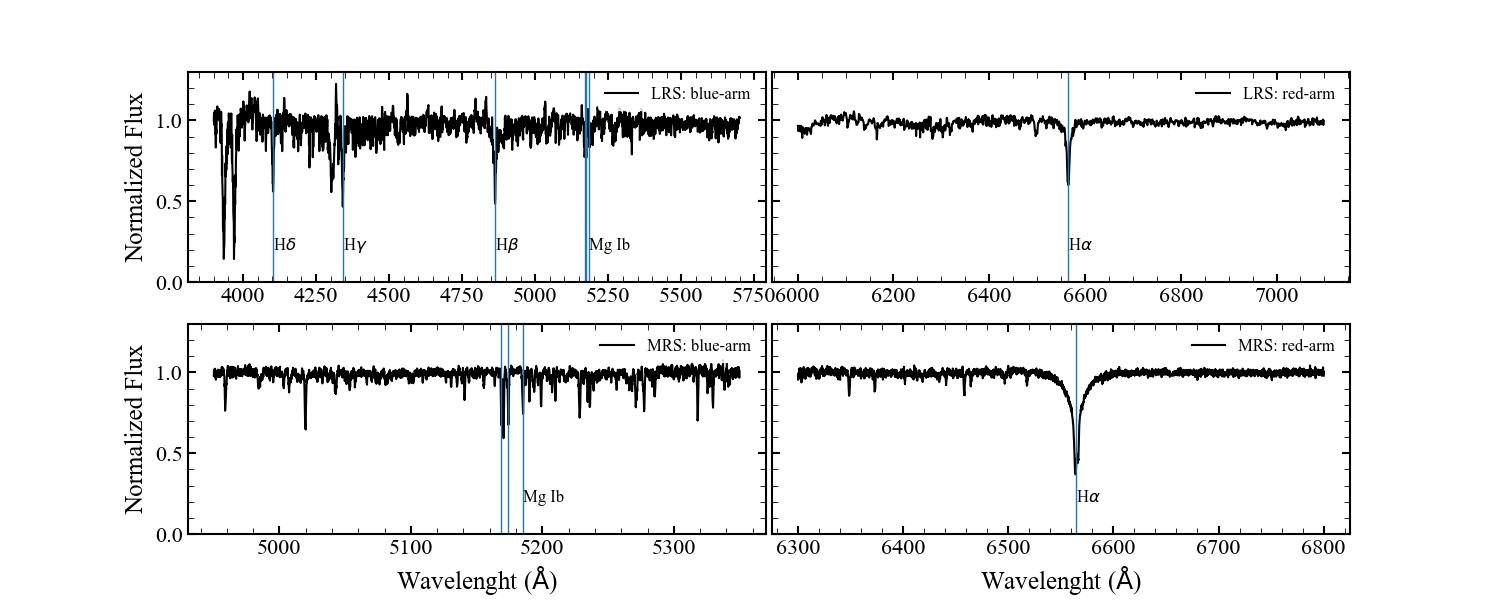}
\caption{
The single exposure LRS and MRS of RRab EPIC\,212182292.
The top two panels illustrate the segments from the red and blue arms of the LRS, and the bottom two panels depict those from the MRS. 
The Balmer lines and Mg\,Ib triplets of LRS and MRS are highlighted with cyan lines.
\label{fig:spec_show}}
\end{figure*}

The LRS and MRS are processed using LASPEC pipeline \citep{2021ApJS..256...14Z}, which includes the removal of cosmic rays, normalization, and radial velocities (\rv) determination.
The continuous of normalization is obtained by least-squares fitting of a 6th-order polynomial.
Figure\,\ref{fig:spec_show} displays the blue and red arms of the normalized LRS and MRS for RRab type EZ Cnc, the SNR of the spectra are 100 and 50, respectively.
The edge bands of blue- and red-arm for LRS are not shown because of the high noise.
The Balmer series and Mg\,Ib triplet lines of the LRS and MRS are labeled with cyan lines and text.

In order to compare directly with the template, we determine the \rv\ of LRS and MRS, respectively, through the cross-correlation function (CCF) in the LASPEC pipeline.
Duo to the intrinsic large amplitude expansion and contraction of RRLs, stellar atmospheres display non-synchronous motions. 
As a result, the velocities of lines originating from various atmospheric altitudes differ significantly \citep{2019A&A...623A.109G}.
Consequently, we derive the \rv\ of both Balmer and metal lines separately.
For LRS, we have determined \rvlalpha, \rvlbeta, \rvlgamma, \rvldelta, and \rvlmetal, which represent the \rv\ of \halpha, \hbeta, \hgamma, and metal lines, from the segments of \slalpha\, ($6450\leq \lambda \leq6700$\,\AA), \slbeta\,($4700\leq\lambda\leq5000$\,\AA), \slgamma\,($4250\leq\lambda\leq4400$\,\AA), \sldelta\,($4000\leq\lambda\leq4250$\,\AA), and \slm\,($5000\leq\lambda\leq5700$\,\AA), respectively.
For MRS, we have determined \rvmalpha\ and \rvmm, which represent the \rv\ of \halpha\ and metal lines, from \smalpha\ ($6500\leq \lambda \leq6650$\,\AA) and \smb\ ($5000\leq\lambda\leq5300$\,\AA) segements, respectively.
The RRLs are generally metal-poor, so the strong lines of the Mg\,Ib triplets primarily contribute to the \rvmetal\ determined from the \slm\ of LRS and the \smb\ of MRS, as illustrated in Figure\,\ref{fig:spec_show}.

\subsection{Stellar atmospheric parameters}
\label{subsec:flow}

In order to weaken the influence of pulsation, we determine the stellar atmospheric parameters of RRLs using an improved template matching method, described as follows:

\begin{itemize}
\item 
Select segments

The Balmer lines are very sensitive to variations in temperature, and have little dependence on variations in \logg, making them a good indicator of \teff.
The Mg\,Ib triplets, covered by LRS and MRS, are very sensitive to variations in \logg, and their profiles depend little on pulsations. 
Therefore, the \teff, \logg, and \mh\ are determined by the Balmer series, Mg\,Ib triplet, and metal lines, respectively.
For LRS, we determine \teff\ from the \slalpha, \slbeta, \slgamma, and \sldelta, while \logg\ and \mh\ from the \slm. 
For MRS, we estimate \teff\ from the \smr\, ($6300\leq\lambda\leq6800$\,\AA), while \logg\ and \mh\ from the \smb.

\item
Maske the line cores of the Balmer series

The Kurucz stellar atmospheric model is not considered the chromosphere, leading to inadequate line core fitting. 
Additionally, line profiles are easily distorted near the phase of peak luminosity, for instance the emission and doubling of the \halpha\ \citep{2014A&A...565A..73G, 2021MNRAS.500.2554B}. 
Therefore, the core of Balmer series lines obtained by Gaussian fitting below the half height is masked.

\item
Initial and final stellar atmospheric parameters

In order to avoid the influence of \rv\ difference between different lines, we shift the Balmer series and metal segments in the velocity space of their corresponding lines before directly comparing them with the templates.
The difference between the observed and synthetic spectra is described by the sum of squares of comparing residuals \chitwo.
\chitwo\ is defined as:
\begin{equation} \label{eq:chi2}
\chi^2 = \frac{1}{N-1}\sum^{N}_{i=1}\frac{(O_i-T_i)^2}{\sigma^2_i}
\end{equation}
where $O_i$ and $T_i$ are the flux of the target and synthetic spectra at the $i$th data point, respectively. $\sigma_i$ and $N$ are the flux error of the $i$th data point and the total number of data points, respectively.
The determination of initial stellar atmospheric parameters is first to determine \teff\ through the segment of Balmer lines, and then to determine \logg\ and \mh\ through the segment of metal lines after fixing \teff.
The templates are then focused around the triple grid interval of the initial stellar atmospheric parameters to simultaneously determine the final \teff, \logg, and \mh.
Each parameter is derived using a weighted average based on an optimal template selected by $\chi^2 \leq a\cdot \chi^2_\mathrm{min}$.
Where $\chi^2_\mathrm{min}$ is the minimum \chitwo\ and $a = 1.01$ is a constant chosen to strike a balance between computational efficiency and accuracy.
The weights are calculated using the formula $w = 1 - \frac{\chi^2 - \chi^2_{\rm min}}{\chi^2_{\rm max}-\chi^2}\times 0.5$, where, $\chi^2_{\rm max}$ is the maximum \chitwo\ in the optimal model \citep{2015MNRAS.448..822X}.
\end{itemize}

We determined the \rv\ of different lines and the stellar atmospheric parameters for LRS and MRS, which are provided in Table\,\ref{tab:lsap} and \ref{tab:msap}.
The full tables can be downloaded at the LAMOST DR10 value-added catalogs website\footnote{www.lamost.org}.

\begin{longrotatetable}
\begin{table}[htbp] 
\startlongtable
\begin{deluxetable*}{llllllllllllllllllllll}
\label{tab:lsap}
\tablecaption{The radial velocities and stellar atmospheric parameters of LAMOST LRS for 10,486 RRLs.
}
\centering
\tablehead{
\colhead{$Gaiaid$} &
\colhead{obsid} &
\colhead{R.A.} &
\colhead{DEC.} &
\colhead{period} &
\colhead{$\varphi$} &
\colhead{Type} &
\colhead{HJD} &
\colhead{SNR$_\mathrm{g}$} &
\colhead{\rvlmetal} \\
\colhead{} &
\colhead{} &
\colhead{} &
\colhead{} &
\colhead{day} &
\colhead{} &
\colhead{} &
\colhead{$-2450000.5$} &
\colhead{\kms} \\
\colhead{(1)} &
\colhead{(2)} &
\colhead{(3)} &
\colhead{(4)} &
\colhead{(5)} &
\colhead{(6)} &
\colhead{(7)} &
\colhead{(8)} &
\colhead{(9)} &
\colhead{(10)} \\
\hline
\colhead{\rvlalpha} &
\colhead{\rvlbeta} &
\colhead{\rvlgamma} &
\colhead{\rvldelta} &
\colhead{\teff $^a$} &
\colhead{\logg $^a$} &
\colhead{\mh $^a$} &
\colhead{Ref.} &
\colhead{flag} \\
\colhead{\kms} &
\colhead{\kms} &
\colhead{\kms} &
\colhead{\kms} &
\colhead{K} &
\colhead{dex} &
\colhead{dex} &
\colhead{} &
\colhead{} \\
\colhead{(11)} &
\colhead{(12)} &
\colhead{(13)} &
\colhead{(14)} &
\colhead{(15)} &
\colhead{(16)} &
\colhead{(17)} &
\colhead{(18)} &
\colhead{(19)} \\
}
\startdata
316067369162854528 & 159108049 & 01:38:06.34 & +33:00:02.41 & 0.652606 & 0.66 & RRab & 6560.7535 & 36.76 & -109$\pm$3 \\
-89$\pm$2 & -103$\pm$2 & -110$\pm$3 & -120$\pm$4 & 6221.0$\pm$98 & 2.37$\pm$0.48 & -2.71$\pm$0.13 & $Gaia$ & [0.99,-3.72,0,0] \\
3694931747383350272 & 203012066 & 12:20:28.08 & -01:33:13.73 & 0.592236 & 0.57 & RRab & 6659.8917 & 26.16 & 170$\pm$3 \\
187$\pm$3 & 170$\pm$2 & 171$\pm$2 & 150$\pm$6 & 6369.0$\pm$95 & 2.41$\pm$0.50 & -1.00$\pm$0.11 & $Gaia$ & [0,0,0.99,0] \\
4006725443696129024 & 508001157 & 12:09:49.73 & +27:33:13.37 & 0.343563 & 0.82 & RRc & 7756.9049 & 10.24 & -103$\pm$33 \\
-197$\pm$7 & -171$\pm$4 & -176$\pm$14 & -156$\pm$10 & 7286.0$\pm$161 & 3.47$\pm$0.43 & -0.58$\pm$0.24 & $Gaia$ & [0,-10.51,0.93,0] \\ 
... & ... & ... & ... & ... & ... & ... & ... & ... & ... \\
... & ... & ... & ... &... & ... & ... & ... & ... \\
\enddata
\vspace{\topsep}
\tablecomments{
This table is available in its entirety in machine-readable form.
$^a$ The error represents the standard deviation of the parameters of the optimal template obtained in template matching.
The description of our catalog：
(1) $Gaiaid$: The $Gaia$ source names from $Gaia$ DR3 RRLs catalog;
(2) obsid: The unique identification number ID of the LAMOST spectrum;
(3) R.A.: The input right ascension (epoch J2000.0) to which the fiber was pointed (in hh:mm:ss.ss);
(4) Dec.: The input declination (epoch J2000.0) to which the fiber was pointed (in dd:mm:ss.ss);
(5) Period: The period of RRLs comes from the catalog in Ref.;
(6) $\varphi$: The phases derived from the $t_0$ and period provided through the catalog in {\tt\string Ref.};
(7) Type: The classifications of RRLs provided by the catalog of Ref.;
(8) HJD: The Heliocentric Julian Date (HJD -2450000.5) at mid-exposure;
(9) SNR$_g$: The signal-to-noise ratio (SNR) of the LRS in the SDSS g band, which is an indicator for the quality of the spectrum;
(10) \rvlm: The radial velocities of metal lines determined from metal segment (\slm);
(11) \rvlalpha: The radial velocities of \halpha\ line determined from \halpha\ segment (\slalpha);
(12) \rvlbeta: The radial velocities of \hbeta\ line determined from \hbeta\ segment (\slbeta);
(13) \rvlgamma: The radial velocities of \hgamma\ line determined from \hgamma\ segment (\slgamma);
(14) \rvldelta: The radial velocities of \hdelta\ line determined from \hdelta\ segment (\sldelta);
(15) \teff: The effective temperature;
(16) \logg: The surface gravity;
(17) \mh: The metallicity;
(18) Ref.: The catalog in which the period and $t_0$ are obtained, including $Gaia$: \citet{2022arXiv220606278C}, ASAS-SN: \citet{2023MNRAS.519.5271C}, ZTF: \citet{2020ApJS..249...18C}, PS1: \citet{2017AJ....153..204S};
(19) flag: A list containing the values of four parameters.
The first three values denote the probabilities of RRLs provided by ASAS-SN, ZTF, and PS1 in the respective references, where a value of 0 signifies that the target is not included in the reference catalog. 
The last value indicates the count of contaminated targets within the $\Delta d \leq 3.7$ arcsec.
}
\end{deluxetable*}
\end{table}
\end{longrotatetable}

\begin{table}[htbp] 
\startlongtable
\begin{deluxetable*}{llllllllllllllll}
\label{tab:msap}
\tablecaption{The radial velocities and stellar atmospheric parameters of LAMOST MRS for 1,027 RRLs.
}
\centering
\tablehead{
\colhead{$Gaia$ id} &
\colhead{obsid} &
\colhead{R.A.} &
\colhead{Dec.} &
\colhead{period} &
\colhead{$\varphi$} &
\colhead{Type} &
\colhead{HJD} \\
\colhead{} &
\colhead{} &
\colhead{} &
\colhead{} &
\colhead{day} &
\colhead{} &
\colhead{} &
\colhead{$-2450000.5$} \\
\colhead{(1)} &
\colhead{(2)} &
\colhead{(3)} &
\colhead{(4)} &
\colhead{(5)} &
\colhead{(6)} &
\colhead{(7)} &
\colhead{(8)} \\
\hline
\colhead{SNR$_\mathrm{b}$} &
\colhead{\rvmm} &
\colhead{\rvmalpha} &
\colhead{\teff} &
\colhead{\logg} &
\colhead{\mh} &
\colhead{Ref.} &
\colhead{Flag} \\
\colhead{} &
\colhead{\kms} &
\colhead{\kms} &
\colhead{K} &
\colhead{dex} &
\colhead{dex} &
\colhead{} &
\colhead{} \\
\colhead{(9)} &
\colhead{(10)} &
\colhead{(11)} &
\colhead{(12)} &
\colhead{(13)} &
\colhead{(14)} &
\colhead{(15)} &
\colhead{(16)}\\
}
\startdata
1376147275855922176 & 84338052 & 15:43:10.94 & +36:53:33.97 & 0.51198 & 0.46 & RRab & 8567.7684 \\
31.31 & -58.8$\pm$0.2 & -66.1$\pm$0.4 & 6040.0$\pm$89 & 2.62$\pm$0.65 & -1.20$\pm$0.10 &  $Gaia$ & [0,-45.71,0,0] \\
675464975453569664 & 84803006 & 07:54:08.77 & +24:10:42.63 & 0.53255 & 0.94 & RRab & 8890.6552 \\
13.46 & -122.6$\pm$1.2 & -132.0$\pm$2.0 & 7445.0$\pm$50 & 3.23$\pm$0.17 & -1.24$\pm$0.09 &  $Gaia$ & [0.99,0,0.75,0] \\
... & ... & ... & ... & ... & ... & ... & ... \\ 
... & ... & ... & ... & ... & ... & ... &  ... \\
\enddata
\vspace{\topsep}
\tablecomments{
This table is available in its entirety in machine-readable form. 
}
\end{deluxetable*}
\end{table}

\section{Uncertainty analysis}
\label{sec:error}

\subsection{Internal uncertainty}
\label{subsec:merror}

Due to the intrinsic variability of the physical parameters of the RRLs, the internal uncertainty ($\sigma_\mathrm{in}$) of these parameters is not estimated by comparing results from multiple observations of targets. 
Therefore, we estimate the $\sigma_\mathrm{in}$ of the \rv\ using the Mentor-Carlo error calculated by LASPEC \citep{2021ApJS..256...14Z}.
Furthermore, we introduce artificial noise to the high-SNR spectra to the star with multiple observations at the same phase. 
We then compare the parameters determined from the spectra with different levels of artificial noise to estimate the $\sigma_\mathrm{in}$ of the stellar atmospheric parameters.

\begin{figure}
    \centering
    \includegraphics[width=.41\textwidth]{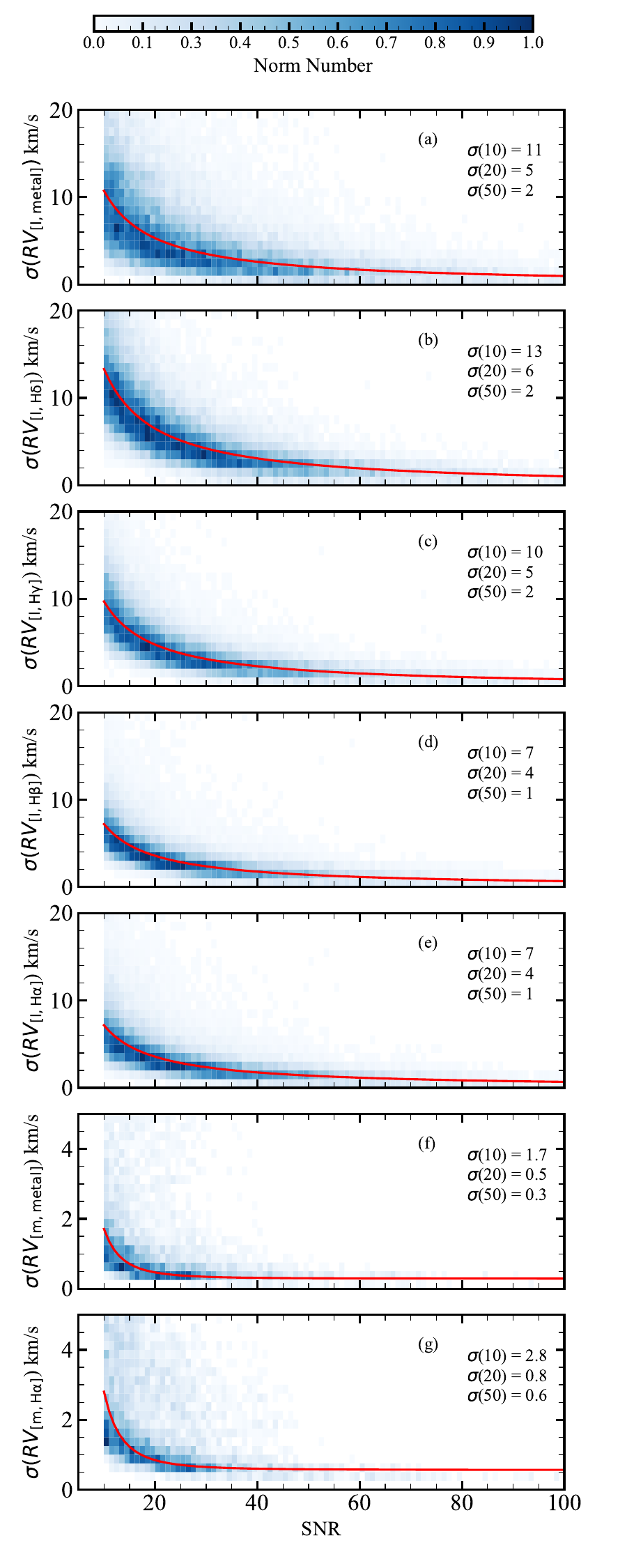}
    \caption{The internal uncertainty of radial velocities for LRS and MRS as a function of SNR. 
    The color represents the normalized number of parameters in the boxes.
    The panels (a)--(e) are the internal uncertainty of \rvlmetal, \rvldelta, \rvlgamma , \rvlbeta, and \rvlalpha\ determined from \slm, \sldelta, \slgamma, \slbeta, and \slalpha, respectively.
    The panels (f) and (g) are the internal uncertainty of \rvmm\ and \rvmalpha\ determined from \smb\ and \smalpha, respectively.
    In each panel, the red line indicates the median within bins with a SNR of 5.
    The accompanying text in each panel specifies the $\sigma$ for \rv\ SNR values of 10, 20, and 50, respectively.
    \label{fig:rv_snerr}}
\end{figure}

The $\sigma_\mathrm{in}$ of \rv\ ($\sigma$(\rv)$_\mathrm{in}$) as a function of SNR is illustrated in Figure \ref{fig:rv_snerr}.
As predicted, the $\sigma$(\rv)$_\mathrm{in}$ decreases with the increase of SNR.
The $\sigma_\mathrm{in}$ of \rvlmetal, \rvldelta, \rvlgamma, \rvldelta, and \rvlalpha\ are 11, 13, 10, 7, and 7\,km/s, respectively, as indicated by the fitted red curves at a SNR$_g\sim$ 10 shown in Figure \ref{fig:rv_snerr}\,(a)--(e).
When SNR$_g$ $\geq$ 20, the $\sigma_\mathrm{in}$ of these \rv\ decreases to 5, 6, 5, 4, and 4\,km/s.
In Figure\,\ref{fig:rv_snerr}\,(f) and (g), the $\sigma _\mathrm{in}$ of \rvmm\ and \rvmalpha\ are 1.7 and 2.8\,km/s, respectively, for a SNR$_b$ of 10.
When SNR$_b\geq$ 50, their $\sigma$(\rv)$_\mathrm{in}$ decrease to 0.3 and 0.6\,km/s, respectively, and remains nearly constant.
This result is consistent with that of \citet{2019RAA....19...75L}, who found a precision of 0.3\,\kms\ for \rv\ at SNR $\sim$ 50.

\begin{figure*}[htbp]
    \centering
    \includegraphics[width=\textwidth]{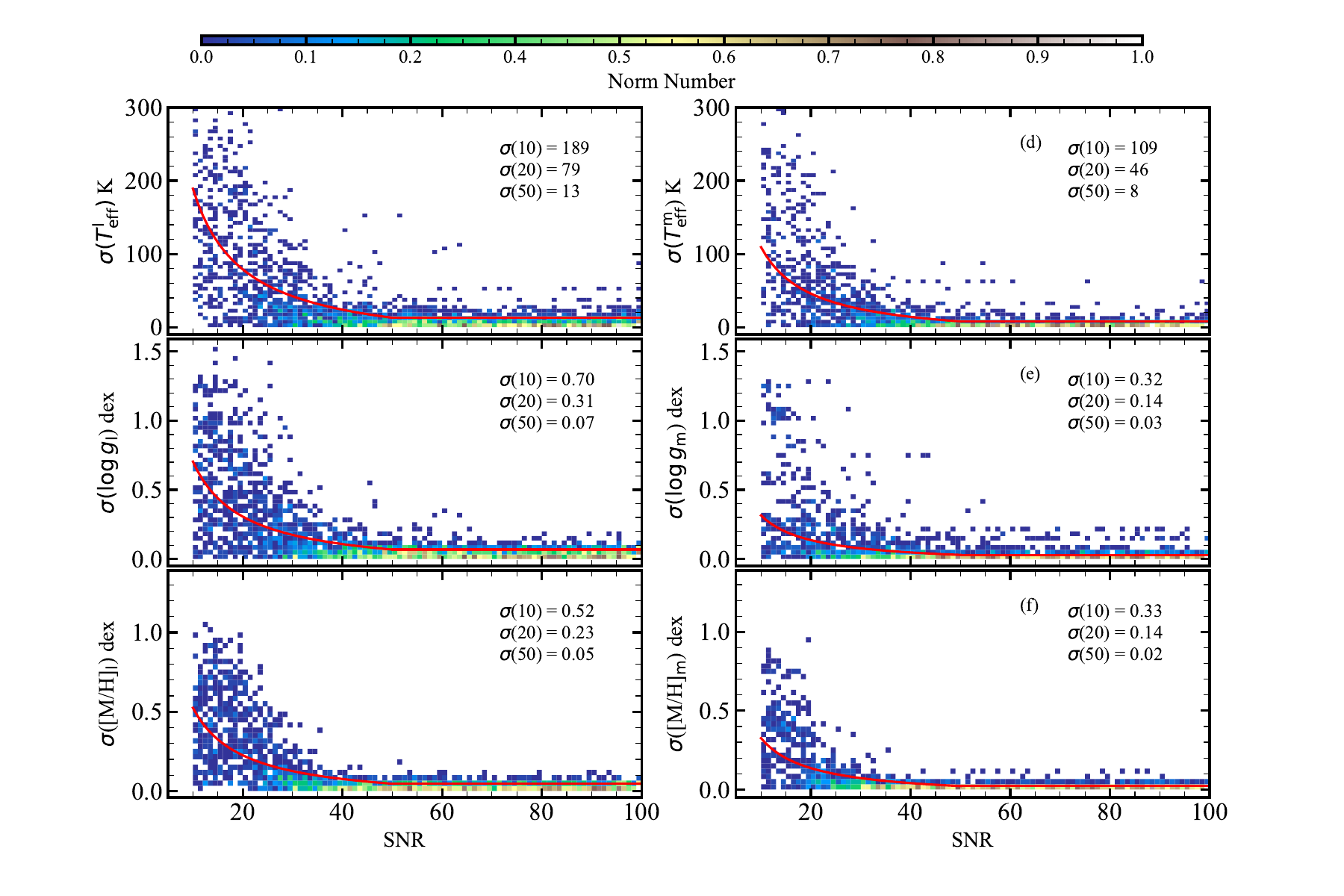}
    \caption{
    The internal uncertainty of \teff, \logg, and \mh\ is determined from LRS ( left panels (a)-(c)) and MRS (right panels (d)-(f)) as a function of SNR.
    The color represents the normalized number of parameters in the boxes.
    The red curve represents the median of the best fit as a reciprocal function of the SNR, and its values at SNR = 10, 20, and 50 are given throughout the text.
    \label{fig:sap_snerr}}
\end{figure*}

We randomly add noise to the 134 LRS and 76 MRS selected based on a criterion with a SNR higher than 100 and determine their stellar atmospheric parameters. 
The internal uncertainty is derived using the equation $\sigma_\mathrm{in}^i(P) = |(P_i - \bar{P}|$, where $i = 1, 2, ..., n$ represents the $i$th set of values of the parameter $P_i$, $n=20$ denotes the total number of added noise sets for the same LRS/MRS, and $\bar{P}$ is the weighted (SNR) average of $n$ sets of $P$.
Figure \ref{fig:sap_snerr} present the  $\sigma_\mathrm{in}$ of \teff, \logg, and \mh\ as a function of SNR.
The $\sigma_\mathrm{in}$ of stellar atmospheric parameters for LRS and MRS decrease with the increase of SNR.
The red line represents the fitted curve obtained from the reciprocal function of $\sigma_\mathrm{in} = \frac{a}{\mathrm{SNR}} + b$, and the corresponding fitted coefficients are presented in Table\,\ref{tab:fit}.
The fitting results are valid with SNR $\leq 50$, because the $\sigma_\mathrm{in}$ is constant when the SNR is higher than 50.
For SNR $\sim$ 10, the internal uncertainty of \teff, \logg, and \mh\ for LRS/MRS are 189/109\,K, 0.70/0.32\,dex, and 0.52/0.33\,dex, respectively, and they reach a stable level of 13/8\,K, 0.07/0.03\,dex, and 0.05/0.02\,dex when SNR $\geq50$.

\begin{table}[htbp] 
\startlongtable  
\label{tab:fit}
\begin{deluxetable}{lrrrrr}
\tablecaption{The coefficients of the optimal fit to estimate the $\sigma_\mathrm{in}$ for the stellar atmospheric parameters.
}
\tablehead{
{} & \multicolumn{2}{c}{LRS} & {} & \multicolumn{2}{c}{MRS} \\
\colhead{} &
\colhead{$a$} &
\colhead{$b$} &
{} &
\colhead{$a$} &
\colhead{$b$}
}
\startdata
\teff & 2202.8 & -31.2 &  & 1269.2 & -17.7 \\
\logg & 7.9   & -0.1  &  & 3.6    & 0.0    \\
\mh   & 6.0   & -0.1  &  & 3.8    & 0.0    \\
\enddata
\end{deluxetable}
\end{table}

\subsection{Pulsation variation}
\label{subsec:pulsa}

To investigate the pulsation characteristic of stellar physical parameters for RRLs, we only select those targets with at least three measurements of stellar atmospheric parameters and an SNR $\geq$ 30 (with $g$ band for LRS or blue-arm for MRS).
The phase of RRLs sample is derived from the spectral observation times (HJD), periods ($P$), and epoch zero ($t_0$), where the $t_0$ is defined as at maximum luminosity.

\begin{figure*}
\centering
\includegraphics[width=\textwidth]{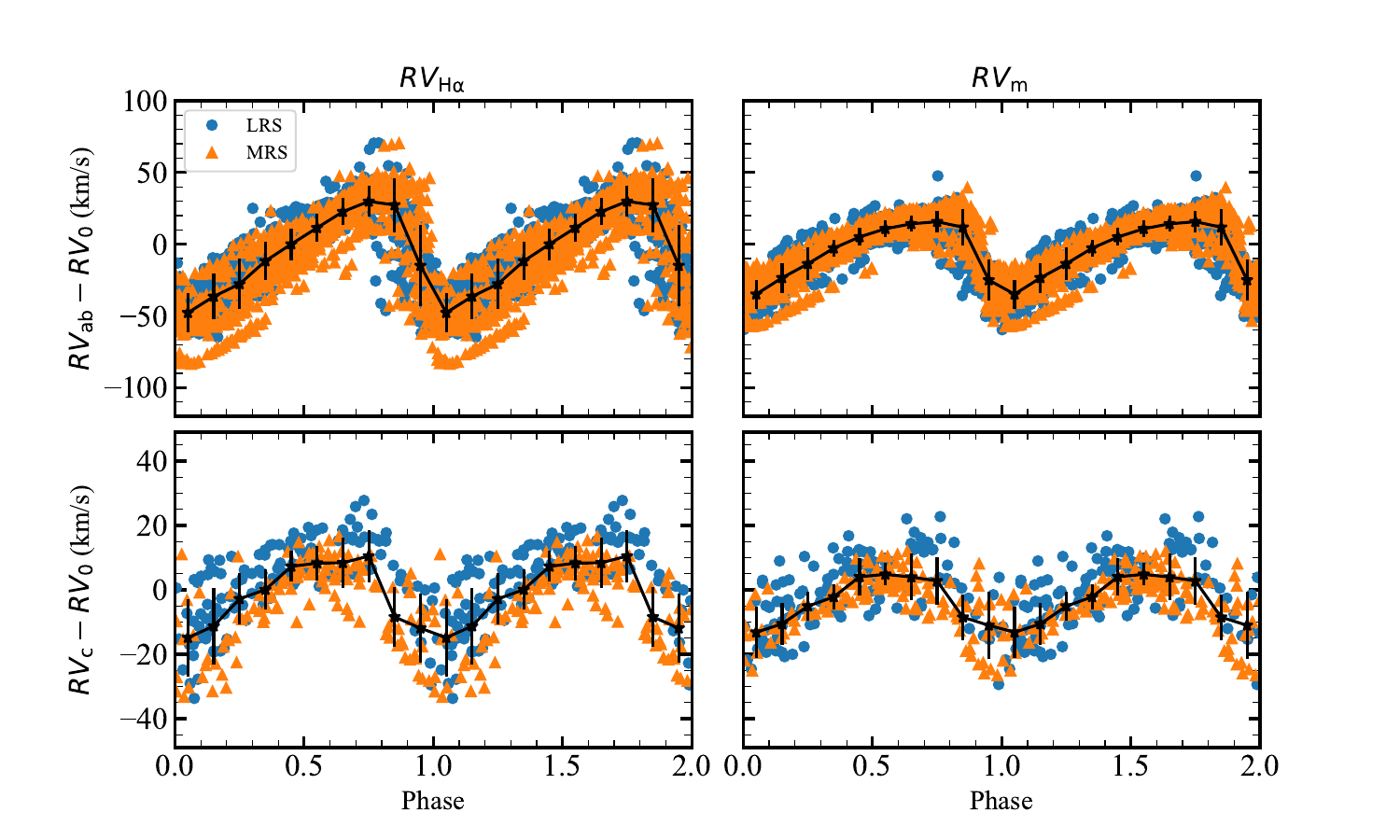}
\caption{The variations of radial velocities during the pulsation cycle.
Where the left panels represent the \rvalpha\, while the right panels are for the \rvmetal, and the top two panels show the \rv\ of RRab ($RV_\mathrm{ab}$), while the bottom panels are for RRc ($RV_\mathrm{c}$).
The blue and orange data correspond to LRS and MRS, respectively.
The black line is the mean within the phase of 0.1 bins, and the error bars indicate the standard deviation.
The center of mass radial velocity ($RV_0$) is obtained from the RRLs catalog of $Gaia$ DR3 \citep{2022arXiv220606278C}.
\label{fig:rvar}}
\end{figure*}

Figure \ref{fig:rvar} shows the phase folding diagrams of \rvalpha\ and \rvmetal\ for RRab and RRc.
These \rv s have been corrected using the radial velocity of the center of mass (\rv$_0$) from the $Gaia$ RRLs catalog \citep{2022arXiv220606278C}.
Within the dataset, there are 121 RRab and 25 RRc in the LRS, and 98 RRab and 21 RRc in the MRS.
The black line is the average value of \rv s at phases with 0.1 bins, and the error bars represent the 99\%\ confidence interval.
In the top panel\,(a) and (b), it can be seen that the average \aptp\ value of \rvalpha\ is $78\pm17$ \kms\ for RRab, which is larger than $50\pm13$ \kms\ of \rvmetal.
For RRc, the \aptp\ of \rvalpha\ and \rvmetal\ are $26\pm13$ and $22\pm11$ \kms, respectively (the bottom panel (c) and (d)).
We find that the \aptp\ of \rv\ for RRc is smaller than that for RRab, because the \aptp\ of \rv\ is positively related to the amplitude of light curves, and the RRab type stars have a larger average \aptp\ of light curves than that of RRc \citep{2021ApJ...919...85B}.

\begin{figure*}[h!]
\centering
\includegraphics[width=\textwidth]{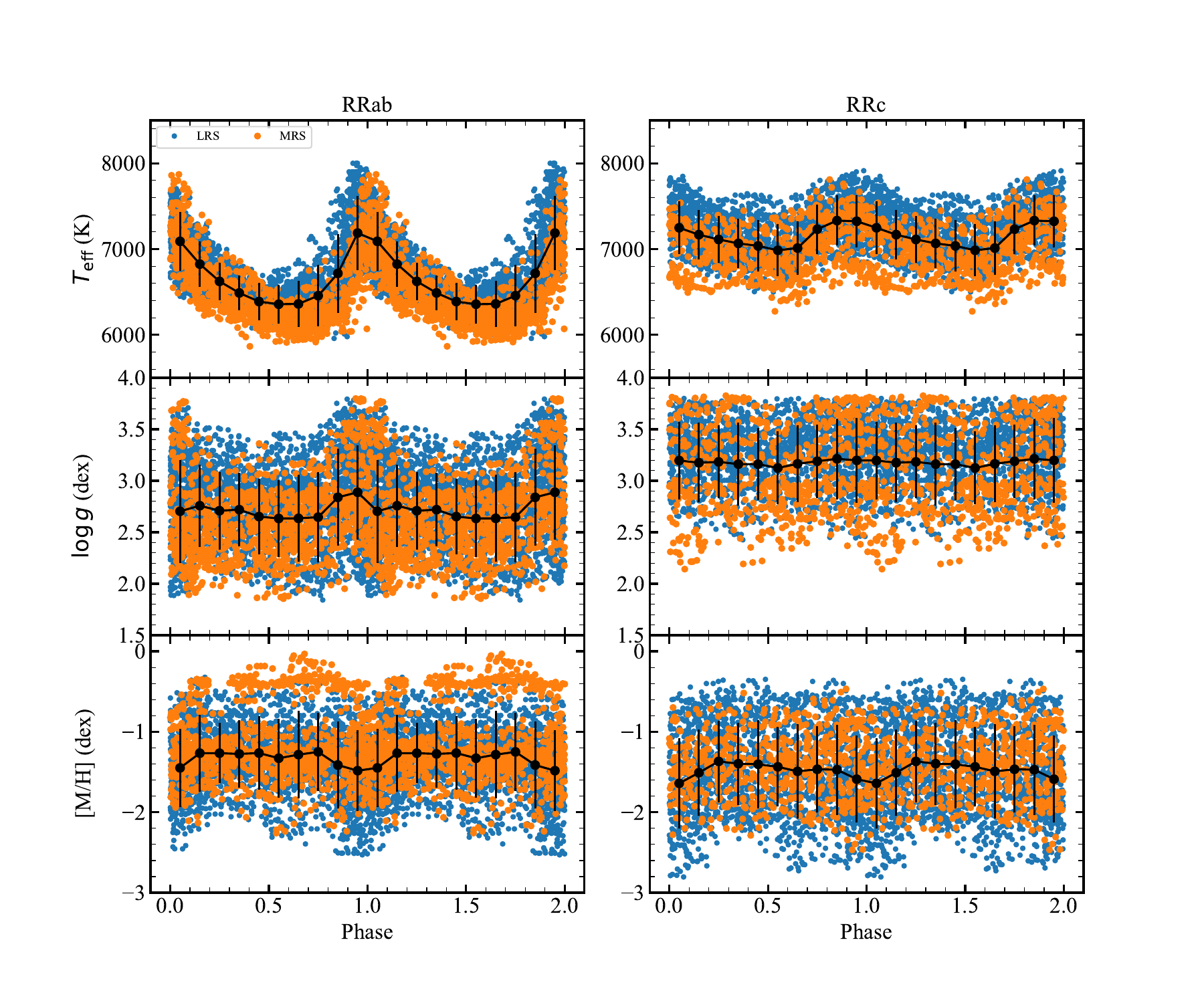}
\caption{The variations of stellar atmospheric parameters during the pulsation cycle.
The left panels represent the LRS, while the right panels are for MRS.
From the top to the bottom panels, the variation in \teff, \logg, and \mh\ with phase are shown.
The method for determining mean values curves (black line) and error bars is same with those employed in Figure \ref{fig:rvar}.
\label{fig:vara}}
\end{figure*}

Figure \ref{fig:vara} illustrates the variation of stellar atmospheric parameters as a function of phase. 
In the top panel, the variation of \teff\ during the pulsation cycle is significant.
According to the average value curves, for RRab stars, the variation of \teff\ ranges from $6264\pm171$ to $7194\pm423$\,K, with an average \aptp\ of $930\pm456$\,K. 
For RRc stars, the \teff\ varies between $6924\pm261$ and $7333\pm256$\,K, with the average \aptp\ being $409\pm375$\,K.
In the middle panel, the \logg\ of RRab has a small amplitude variation of about $0.22\pm0.42$\,dex during the pulsation phase, while that of RRc remains almost constant.
In the lower panel, the \mh\ values of RRab and RRc show slight variations of $0.25\pm0.50$ and $0.28\pm0.55$\,dex, respectively, while the values for the other phases remain almost consistent.
In addition, there are many \mh\ of RRab from MRS distributed near the metal-rich $-0.50$\,dex, and remains nearly constant during the pulsation cycle.
Most of the data points are contributed from {\tt\string Gaia\,DR3\,689686230645254912} (EZ\,Cnc), {\tt\string Gaia\,DR3\,255409438666669824}, and {\tt\string Gaia\,DR3\,468529087451089024}, with 57, 34, and 27 MRS, respectively. 
Specifically, the weighted average value of \mh\ for EZ Cnc is $-0.46\pm0.04$\,dex, which is in good agreement with the metallicity value of $Z=0.006\pm0.002$ from astroseismology \citep{2021MNRAS.506.6117W}.

\subsection{Compare with other databases}
\label{subsec:compar}

To evaluate the external uncertainty ($\sigma_\mathrm{out}$) of the stellar atmospheric parameters, we compare our results with those from other databases, including  $Gaia$, APOGEE, SEGUE, and \citet{2019RAA....19...75L}. 
Considering that the stellar atmospheric parameters provided by the LAMOST DR10\footnote{https://www.lamost.org/dr10/v1.0/doc/release-note} and GALAH DR3 \citep{2021MNRAS.506..150B} lacks coverage of Horizontal Branch stars, we do not compare with them.
Assuming a coordinate matching difference of less than 3.0 arcsecs is more strict than that in Section \ref{subsec:samp}, we establish the following criteria for common targets to minimize dispersion attributable to phase differences.
\begin{itemize}

\item 
SNR limitation: The SNR$_g$ of LRS and SNR$_b$ of MRS are higher than 20.

\item 
Difference of phase constraints:
We define $\Delta RV = \Delta \varphi \cdot A_{RV}^\mathrm{metal}$ to convert the phase difference ($\Delta \varphi$) between common targets into their \rv\ difference ($\Delta RV$) to address the lack of observation times or phases in other databases, where $A_{RV}^\mathrm{metal}$ denotes the \aptp\ of \rvmetal.
To ensure that the phase difference is sufficiently small, while the number of common targets is sufficiently large, we set $\Delta \varphi = 0.2$. 
The corresponding $\Delta RV$ values for RRab and RRc are approximated as 10 and 5\,\kms, respectively.

\item 
Weakening pulsation effects: 
The phase was restricted to a range of $0.2 \leq \varphi \leq 0.8$ to avoid anomalous variations in stellar atmospheric parameters near the 0.9 phase.

\end{itemize}

\begin{figure*}[ht!]
    \centering
    \includegraphics[width=\textwidth]{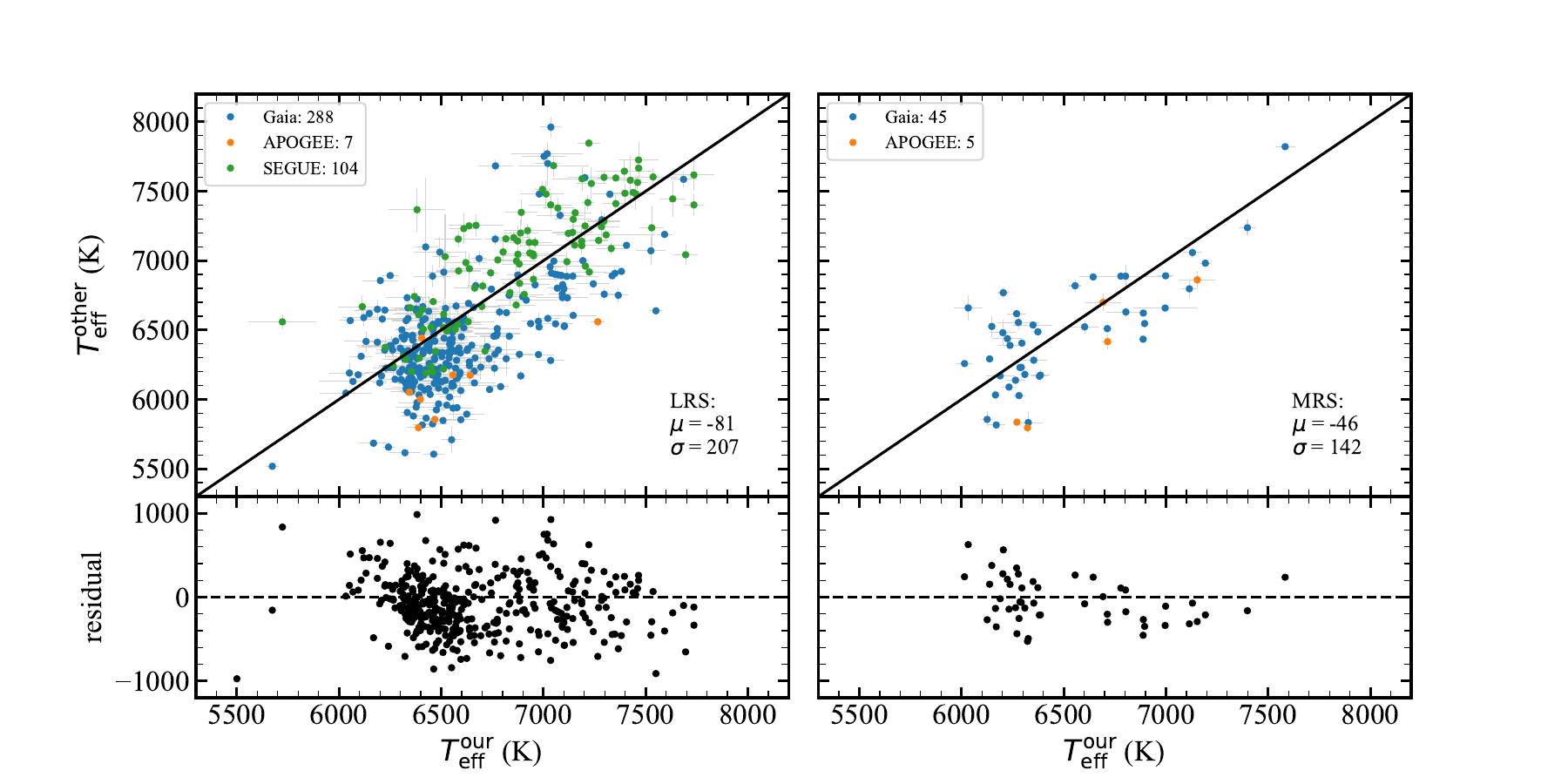}
    \caption{The comparison of \teff\ with the other datasets.
    The blue, orange, and green data correspond to $Gaia$, APOGEE, and SEGUE, respectively, in legend. 
    The numbers indicate the count of common targets.
    The residuals of \teff\ between other datasets and our values are shown at the bottom, along with their mean ($\mu$) and standard deviation ($\sigma$) in the text.
    The black solid and dashed lines indicate the bisector and zero position, respectively.
    The left panel displays \teff\ of LRS, and the right panel for MRS.
    \label{fig:compare_teff}}
\end{figure*}

When compared to the other database, we adopt the weighted average of the stellar atmospheric parameters.
The comparison of \teff\ is shown in Figure \ref{fig:compare_teff}.
In the left panel, the common targets of our LRS sample with $Gaia$, APOGEE, and SEGUE are 288, 7, and 104, respectively.
The data points cluster closely around the bisector line, and the mean ($\mu$) and standard deviation ($\sigma_\mathrm{out}$) values of residuals are $-81$ and 207\,K, respectively, indicating a good consistency between our results and other datasets. 
In the right panel, the \teff\ for MRS shows a good consist with these from $Gaia$ and APOGEE, and the $\mu$ and $\sigma_\mathrm{out}$ values are -46 and 142\,K, respectively.

\begin{figure*}[ht!]
    \centering
    \includegraphics[width=\textwidth]{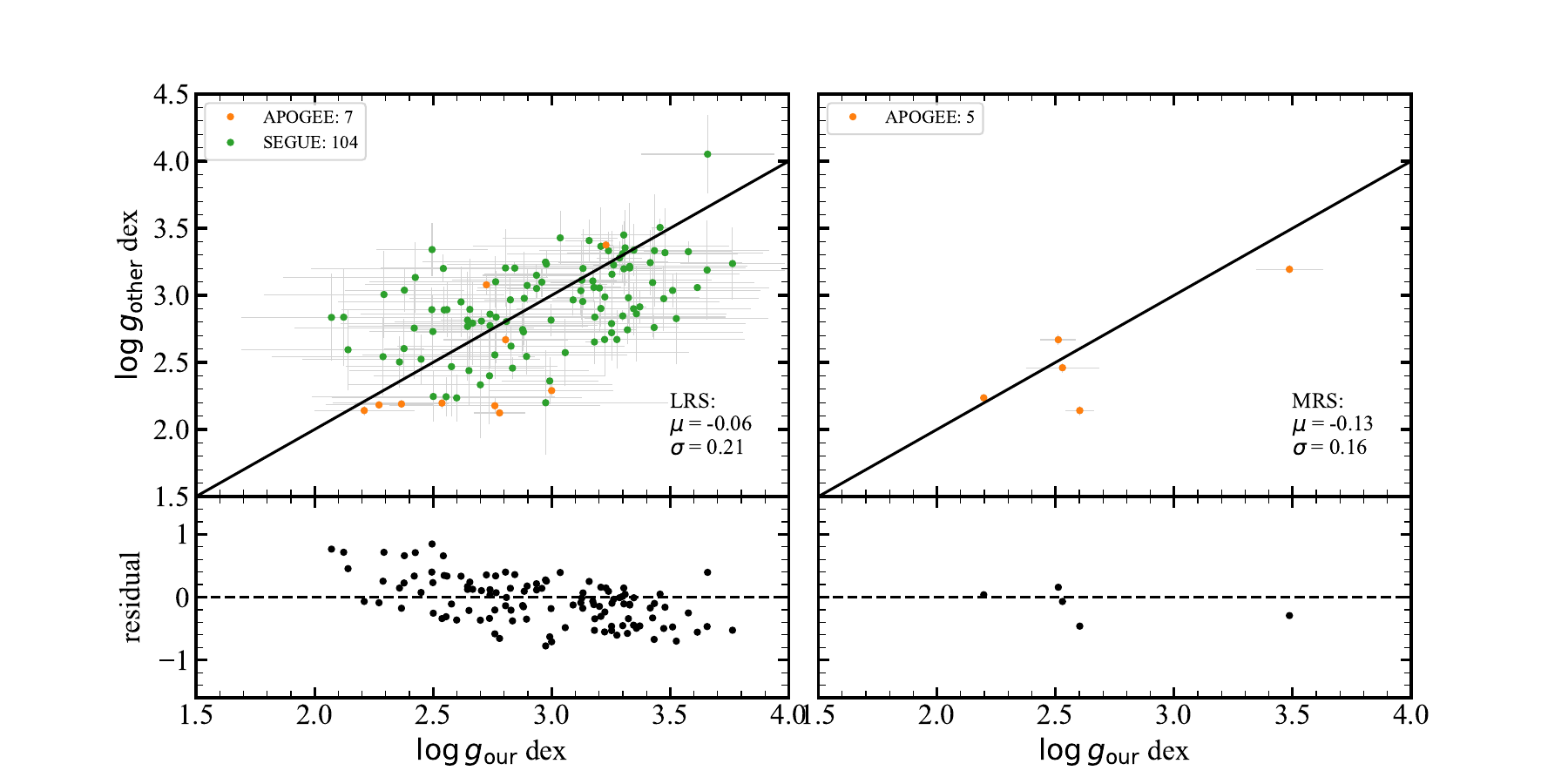}
    \caption{The comparison of \logg\ with the other datasets.
    The colors of data and legends are the same as in Figure \ref{fig:compare_teff} but for \logg.
    \label{fig:compare_logg}}
\end{figure*}

Figure \ref{fig:compare_logg} presents a comparison of \logg. 
For common targets of LRS, the $\mu$ and $\sigma_\mathrm{out}$ of the residuals are $-0.06$ and 0.21\,dex, respectively.
The comparison of MRS shows a good agreement, with $\mu$ and $\sigma_\mathrm{out}$ of $-0.13$ and 0.16\,dex, respectively. 
The \logg\ may be underestimated for low surface gravity conditions of \logg\,$ < 2.2$\,dex, while overestimated for high surface gravity conditions of \logg\,$ > 3.5$\,dex.

\begin{figure*}[htbp]
\centering
\includegraphics[width=\textwidth]{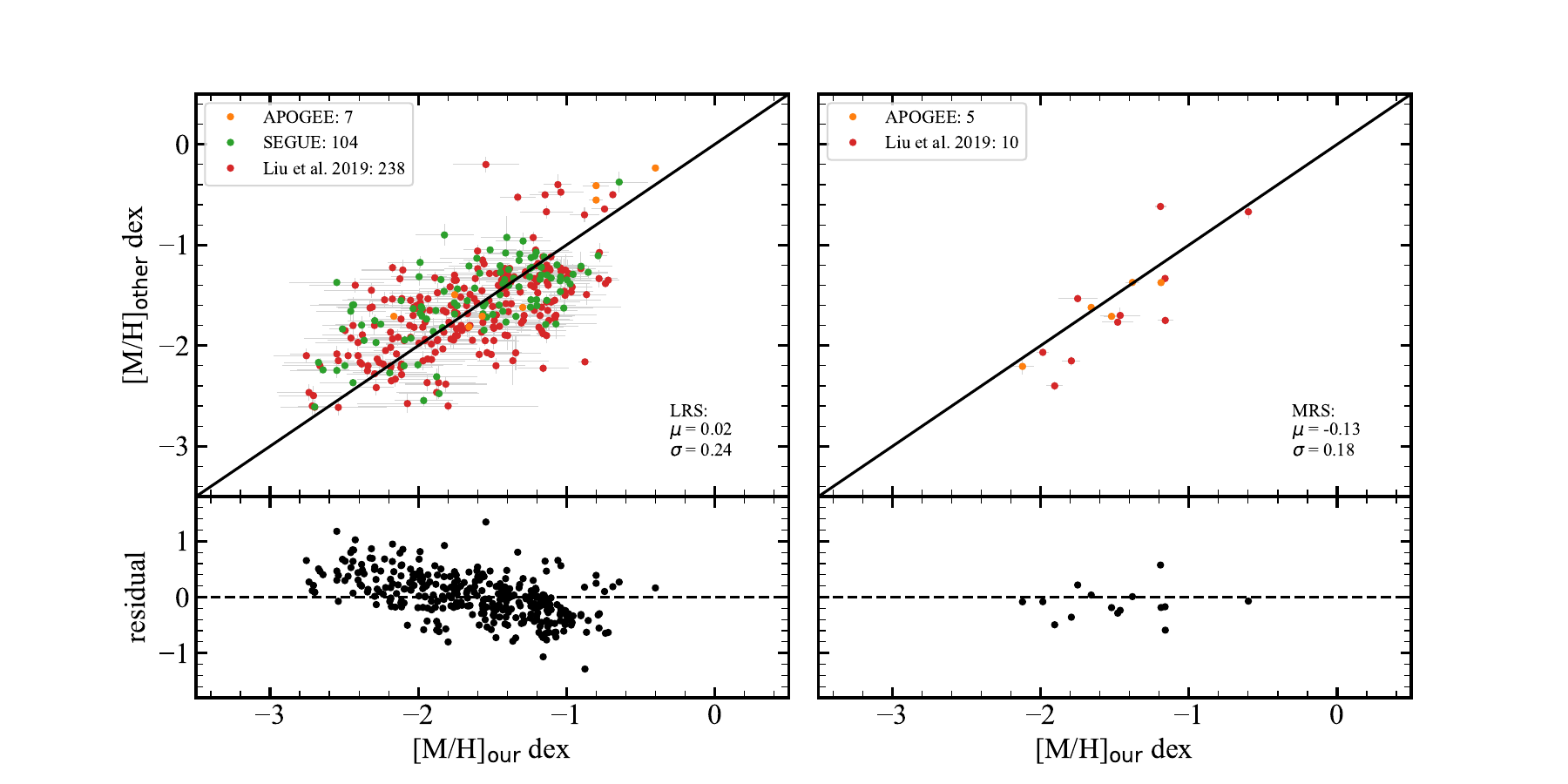}
\caption{The comparison of \mh\ with the other datasets.
        The red points represent the data from \citet{2019RAA....19...75L}, and the other colors and legends are the same as in Figure\,\ref{fig:compare_teff} but for \mh.
\label{fig:compare_mh}}
\end{figure*}

The comparison of \mh\ is illustrated in Figure\,\ref{fig:compare_mh}.
The $\mu$ of residues for LRS and MRS are 0.02 and $-0.13$\,dex, respectively, and $\sigma_\mathrm{out}$ are 0.24 and 0.18\,dex.
For the LRS, the residuals show a linear trend similar to that of \logg,  which indicates that the \mh\ from LRS is underestimated for the metal-poor region of \mh\,$< -2.50$\,dex, while overestimated for the slightly metal-rich region of \mh\,$>-1.00$\,dex.

\section{Discussion}
\label{sec:disc}

In section\,\ref{subsec:pulsa}, we analyze the characteristics of the \rv\ and stellar atmosphere parameters during the pulsation cycle.
The differences of \aptp\ between \rvalpha\ and \rvmetal\ is $28\pm21$\,\kms\ for RRab, while it is $4\pm17$\,\kms\ for RRc.
The differences in \aptp\ of \rv\ between the RRab and RRc from the \halpha\ and metal lines are $52\pm21$ and $28\pm17$\,\kms\, respectively.
Similar differences have been presented in the results of the \rv\ curves of 36 RRLs obtained from the HRS by \citet{2021ApJ...919...85B}.
The difference of the \aptp\ of \rv\ for different lines indicates asynchronous movement within the atmospheres of RRLs at varying altitudes.
This suggests that the Doppler effect of the segments needs to be corrected by the \rv\ of their corresponding lines, which is a key step in the improved template matching method.

There are differences between the stellar atmospheric parameters of the RRab and RRc stars.
For RRab, the \teff\ ranges from $6264\pm171$ to $7194\pm423$\,K with an amplitude of $930\pm456$\,K, while for RRc, it ranges from $6924\pm261$ to $7333\pm256$\,K with an amplitude of $409\pm375$\,K.
The \logg\ of RRab experiences a slight upward and then downward variation around the phase of $\varphi = 0.9$, while that of RRc remains almost constant.
This is because \logg\ represents the effective gravity ($g = g_M + g_r$) resulting from the combination of mass gravity ($g_M = \mathrm{G} M/r^2$) and radius acceleration ($g_r = dRV(t)/dt$) , where $\mathrm{G}$ is the gravitational constant, $t$ represents time, and $M$ and $r$ are the mass and radius of the star, respectively \citep{2013MNRAS.434..552C}. 
As a result, the \logg\ of RRab, which shows larger amplitude compared to RRc, increases more discernibly near the phase of $\varphi = 0.9$ where the $r$ is minimized, and $dRV(t)$ largest \citep{2013MNRAS.434..552C}.
This implies that the position of individual RRLs within the HRD undergoes significant shifts due to the variation of \teff.
In the phase folding diagrams of stellar atmospheric parameters obtained by the HRS for 11 RRab \citep{2011ApJS..197...29F} and 19 RRc stars \citep{2017ApJ...848...68S}, as displayed in \citet{2019ApJ...881..104M}, the characterization of the variations in \teff\ and \logg\ is very similar to our results.
In addition, based on the stellar atmospheric parameters near the phase of $\varphi = 0.5$, we obtain average \teff\ and \logg\ of $6264\pm171$\,K and $2.48\pm0.29$\,dex for RRab, while $6924\pm261$\,K and $2.99\pm0.29$\,dex for RRc, respectively.
This characteristic was already noted by \citet{2019ApJ...881..104M}, indicating that RRab is located closer to the cooler and darker red edge of the IS from \citet{2015ApJ...808...50M} compared to RRc.

According to the statistical results, the amplitude difference between \rvalpha\ and \rvmetal\ in RRab is more than twice that of RRc.
This significant difference suggests that complex movements are more easily generated in the interiors of atmospheres for RRab, which can affect their line profiles.
Consequently, this influences the determination of \logg\ and \mh.
For example, \citet{2019A&A...623A.109G} investigated RR Lyr of RRab type showing the Blazhko effect and concluded that the intensity of shock excitation at the phase of $\varphi \sim 0.90 - 0.943$ is positively correlated with the modulated \aptp\ of light curves,
i.e., the \aptp\ of \rv.
Therefore, in Figure\,\ref{fig:vara}, the \mh\ of RRab and RRc shows a mild decrease of about $0.29\pm0.49$ and $0.26\pm0.52$\,dex near the phase of $\varphi = 0.9$, respectively.

We propose to convert the phase difference constraints between our sample and other databases into radial velocity difference constraints.
The results of this comparison indicate that our stellar atmosphere parameters from MRS show a good agreement with other databases, indicating the reliability of results from MRS.
However, the residuals of the comparisons between our \logg\ and \mh\ from LRS and those from other databases present a linear trend, suggesting that \mh\ and \logg\ from LRS have systematic uncertainties.

\section{summary}
\label{sec:summ}

We collect a sample of about 449,093 RRLs from the catalogs of the $Gaia$, ASAS-SN, ZTF, and PS1 survey projects, and 174,030 RRLs are in the region of the LAMOST field.
Through cross-matching with LAMOST DR10, we obtained 42,729 LRS and 7,064 MRS.

We first determine the \rv s of \halpha, \hbeta, \hgamma, \hdelta, and metal lines for LRS, while the \rv s of \halpha\ and metal lines for MRS.
We then optimize the template-matching method to estimate the stellar atmospheric parameters of the RRLs, as described in Section\,\ref{subsec:flow}.
The spectral information, periods, phase, and determined \rv s and stellar atmospheric parameters of our RRLs sample are presented in Tables \ref{tab:lsap} and \ref{tab:msap}.

We find that the internal uncertainty of \teff, \logg, and \mh\ decreases with increasing SNR, and stabilizes when after SNR $\geq$ 50.
For comparisons with other databases, we only chose stellar atmospheric parameters with the differences of \rvmetal\ less than 10/5\,\kms\ and the phases of 0.2 $\leq \varphi \leq$ 0.8 to avoid uncertainties due to the pulsation. 
We conclude that the systematic differences of \teff, \logg, and \mh\ are $-81/-46$\,K, $-0.06/-0.13$\,dex, and $0.02/-0.13$\,dex for LRS/MRS, respectively, while the $\sigma_\mathrm{out}$ are 207/142\,K, 0.21/0.16\,dex. and 0.24/0.18\, dex.
The final uncertainties ($\sigma$) of the stellar atmospheric parameters are estimated by combining the $\sigma_\mathrm{in}$ and $\sigma_\mathrm{out}$, which are calculated using the formula $\sigma = \sqrt{\sigma_\mathrm{in}^2 + \sigma_\mathrm{out}^2}$.
Where $\sigma_\mathrm{in}$ is a function of SNR at SNR $\leq$ 50, the fitted coefficients are displayed in Table\,\ref{tab:fit}.

We summarize the variation characteristics of the stellar physical parameters of RRLs during the pulsation cycle.
\begin{itemize}
\item
The \aptp\ of \rvalpha\ variation is greater than that of \rvmetal, and this difference is significantly greater for RRab than that for RRc.
The \aptp\ of \rvalpha\ and \rvmetal\ are $78\pm17$ and $50\pm13$\,\kms\ for RRab, and $26\pm13$ and $22\pm11$\,\kms\ for RRc, respectively.
\item 
There is a significant difference in the \teff\ variation of RRab and RRc during the pulsation cycle.
The \teff\ variation ranges from $6264\pm171$ to $7194\pm423$\,K with an \aptp\ of $930\pm456$\,K for RRab, while $6924\pm261$ to $7333\pm256$\,K with an \aptp\ of $409\pm375$\,K for RRc.
\item
The \logg\ of RRab has a small variation near the phase $\varphi = 0.9$, while that of RRc is difficult to identify.
Similar to the \logg\ of RRab, the \mh\ shows a slight variation near the 0.9 phase, with values of $0.25\pm0.50$ and $0.28\pm0.55$\,dex for RRab and RRc, respectively.
\end{itemize}

In conclusion, we present two catalogs of stellar physical parameters, containing about 11,000 RRLs. 
The stellar atmospheric parameters of the large RRLs sample are very helpful for various fields of astronomical research.
According to the metallicity-luminosity relationship, the distance can be determined by the precise metallicities \citep{2018MNRAS.481.1195M}.
The accuracy of the near-infrared period-luminosity relationship can be enhanced through \mh\ calibration \citep{2015ApJ...808...50M, 2021MNRAS.502.4074M, 2023ApJ...944...88L}, which is beneficial for calibrating the zero point of Gaia distances and compensating for the accuracy of Gaia distance for distant targets \citep{2022MNRAS.513..788G}.
Through the variation of \teff\ and \logg\ with a phase, the red and blue edges of the IS can be constrained, and further exploration of the effects of \mh\ on the IS edge can be conducted \citep{2015ApJ...810...15F, 2021ASPC..529..147L}.
In particular, for some RRLs, they have  \rvalpha\ and \rvmetal\ in almost all folding phases, which will help us to understand the kinematic and dynamical information of different envelopes of the internal stellar atmosphere \citep{2019A&A...623A.109G}, and will provide observational conditions on the physical mechanism of the Blazhko effect \citep{2013MNRAS.434..552C}, also will assist in asteroseismology studies in combined with light curves \citep{2021MNRAS.506.6117W}.

\begin{acknowledgments}
Our research is supported by the National Natural Science Foundation of China under grant Nos. 12090040, 12090042, 12090044, 11833006, 11833002, and the National Key R\&D Program of China No. 2019YFA0405502.
JTW acknowledges support from the LAMOST FELLOWSHIP fund.
The LAMOST FELLOWSHIP is supported by Special Funding for Advanced Users, budgeted and administered by the Center for Astronomical Mega-science, Chinese Academy of Sciences (CAMS-CAS).
We acknowledge the science research grants from the China Manned Space Project.
This work is supported by the Cultivation Project for LAMOST Scientific Payoff and Research Achievement of CAMS-CAS.
Guoshoujing Telescope (LAMOST) is a National Major Scientific Project built by CAS. Funding for the project has been provided by the National Development and Reform Commission. LAMOST is operated and managed by the National Astronomical Observatories, CAS.

\end{acknowledgments}

\bibliography{sample631}{}
\bibliographystyle{aasjournal}


\end{CJK*}
\end{document}